# Sustainability and the Astrobiological Perspective:

## Framing Human Futures in a Planetary Context

Adam Frank[1] & Woodruff Sullivan[2]

**Abstract**

We explore how questions related to developing a sustainable human civilization can be cast in terms of astrobiology. In particular we show how ongoing astrobiological studies of the coupled relationship between life, planets and their co-evolution can inform new perspectives and direct new studies in sustainability science. Using the Drake Equation as a vehicle to explore the gamut of astrobiology, we focus on its most import factor for sustainability: the mean lifetime $\bar{L}$ of an ensemble of Species with Energy-Intensive Technology (SWEIT). We then cast the problem into the language of dynamical system theory and introduce the concept of a trajectory bundle for SWEIT evolution and discuss how astrobiological results usefully inform the creation of dynamical equations, their constraints and initial conditions. Three specific examples of how astrobiological considerations can be folded into discussions of sustainability are discussed: (1) concepts of planetary habitability, (2) mass extinctions and their possible relation to the current, so-called Anthropocene epoch, and (3) today's changes in atmospheric chemisty (and the climate change it entails) in the context of pervious epochs of biosphere-driven atmospheric and climate alteration (i.e. the Great Oxidation Event).

[1] Department of Physics and Astronomy, University of Rochester, Rochester NY. afrank@pas.rochester.edu

[2] Dept. of Astronomy and Astrobiology Program, University of Washington, Seattle, WA. woody@astro.washington.edu

# 1. Introduction

Anthropogenic global climate change is currently recognized as a significant, perhaps fundamental, issue facing human civilization (Solomon et al 2007). The chemical composition of the Earth's atmosphere has been significantly altered by human activity. Moreover, detailed analysis of global data sets has implied the potential for driving the climate system into a state quite different from the one in which human civilization has emerged and flourished (Parry et al 2007). Recognition of the likelihood of profound climate change has thus led to the desire to sustain, to some degree, the current climate state. Climate science is, however, just one domain in which discussions of "sustainability" have emerged. It has also gradually become apparent that human activity has been driving many other changes in the coupled "earth systems" of atmosphere, hydrosphere, cryosphere, geosphere and biosphere in ways that could threaten, or at least strongly stress, the so-called "project of civilization."

Such changes to the earth systems include: (a) the depletion of natural fisheries where it is estimated that 95% of all fish stocks have suffered some form of collapse over the last half century (Worm et al 2006); (b) diminishing supplies of fresh water (Gleick 2003); (c) loss of rain forest habitat (Williams 2003); and (d) continuing acidification of the oceans (Cicerone et al 2004). In all cases human activity, integrated over time and location, have led to substantial changes in the state of the coupled earth systems. These changes have been dramatic enough for some researchers to begin speaking of the beginning of the Anthropocene, a new, geological epoch succeeding the Holocene (the current interglacial period; Zalasiewicz et al 2010).

The field of sustainability science has emerged in the wake of this recognition seeking to understand the interactions "between natural and social systems" (Kates 2011). In particular this discipline studies how such interactions can lead to new modalities of human development that meet "the needs of the present and future generations." Sustainability science, bridging disparate domains such as sociology and Earth Systems

Science, has grown rapidly. More than 20,000 articles have appeared addressing sustainability science over the last 40 years with a doubling in the number of articles every 8 years (Kates 2011). Although sustainability science often focuses on place-specific issues, by its very nature it requires a global perspective to address issues associated with the need to "conserve the planet's life support systems" for future generations (Kates 2010). This trend is clear in recent studies exploring planetary-scale tipping points (Lenton & Williams 2013) or the existence of planetary-scale "boundares" as safe operating limits for civilization (Rockstrom et al 2009). In this way sustainability science and the theoretical perspective it takes on the trajectory of human culture is necessarily global, or better yet *planetary*. It is from that perspective that sustainability science overlaps with the domain of another young and rapidly growing field: astrobiology.

Astrobiology is essentially the study of life in an astronomical context (Sullivan & Baross 2007). The NASA Astrobiology Institute, for example, defines its subject as "the study of the origin, evolution, distribution, and future of life in the universe". More specifically, since most extrapolations involve a planetary context for the origin and evolution of life, astrobiology is concerned with planetary issues just like sustainability science.

Astrobiology faces an obvious "N=1 dilemma" in that we have only one known example of life in the universe (more on this in Section 2). Nevertheless, since the 1990s there has been an explosion in new studies and new results relevant to astrobiology's core questions. For the purposes of this paper we break these advances into three research domains (though there are others that are of broader import):

1. Exoplanets: the discovery of planets orbiting other stars and the characterization of exoplanetary systems in terms of habitability for life (Lineweaver & Chopra 2012, Udry 2007).
2. Solar System Studies: the detailed (often *in situ*) exploration of planets, moons and other bodies in our own planetary system with a focus on the evolution and history of habitable locations (i.e., liquid water, sources of free energy for metabolism, etc.) (Lineweaver & Chopra 2012, Arndt & Nisbet 2012).

3. Earth System Studies: the detailed investigation of the Earth's history including the history of the 3.5-Gyr-old biosphere and its coupled interactions with atmosphere, oceans, ice regions and land masses (Azua-Bustos et al 2011, Coustenis & Blac et al 2011).

The most notable discovery in astrobiology relevant to exoplanets (domain 1) has been the recognition that planets are quite common in the Galaxy, with more than a billion Earth-mass-like worlds expected to exist on orbits within the habitable zones of their stars (Seager 2012). Relevant to solar system studies (domain 2) we now recognize that Mars once hosted liquid water on its surface and that many of the moons of the gas and ice giant planets harbor subsurface liquid oceans (Castillo-Rodgez & Lunine 2012). Relevant to earth systems science (domain 3), we realize that the biosphere and non-biological Earth systems have, at least during some epochs, *co-evolved*, meaning that significant feedbacks have led to substantial changes in the evolution of the entire earth system. The development of an oxygen-rich atmosphere due, in part, to the respiration of anaerobic bacteria is one example of such co-evolution (Kasting & Knoll 2012).

Thus astrobiology takes an inherently large-scale and long-term view of the evolution of life and planets. In this way the data, the perspective and conceptual tools of astrobiology may cast the global problems of sustainability science into a different and, perhaps, useful light. In particular, the astrobiological perspective allows the opportunities and crises occurring along the trajectory of human culture to be seen more broadly as, perhaps, critical junctures facing *any species* whose activity reaches significant level of feedback on its host planet (whether Earth or another planet). In this way, the very question of sustainability may be seen not solely through the lens of politics and policy decisions (Kates 2010), but also as an essential evolutionary transformation that all (or at least many) technological species must experience.

In this paper we explore the argument that perspectives developed through astrobiological studies can usefully inform sustainability science by broadening its understanding, providing case studies, and suggesting different modes of conceptualization. In particular, we seek to frame a research program that might allow researchers to develop a better understanding of the types of trajectories a biosphere might follow once a generic Species with Energy-Intensive Technology *(SWEIT)* emerges.

We begin in Section 2 with a discussion of the relevance of astrobiology, using the standard Drake Equation as a vehicle for framing our questions. In addition we use the Drake equation to address the concept of a statistically relevant ensemble SWEITs. We then discuss in Section 3 possible theoretical tools for modeling sustainability from an astrobiological perspective with an emphasis on dynamical system theory. In Section 4 we present three examples of specific astrobiological topics that can inform sustainability: definitions of habitability across time, the occurrence of mass extinctions, and the biosphere driven climate change as a consequence of SWEIT activity. Finally, in Section 5 we summarize our findings as well and discuss possible directions for future research.

## 2. The Drake Equation and its Longevity Factor *L*

Historically the Drake Equation has been instrumental in framing discussions of astrobiology. Originally proposed by Frank Drake in 1962 as a means for estimating the present number (*N*) of radio-transmitting cultures, the equation cleanly parses the question of life and its evolution into astronomical, biological and sociological factors (Tarter 2007). Note that the original intention of the Drake equation was to estimate the number of civilizations detectable today. Since we are interested in the question of SWEIT lifetimes, detectability is not our concern. Instead our first goal is to use equation to estimate the number of SWEITs that exist now or have already gone extinct.

In its traditional form the Drake Equation is

$$N = R_* f_p n_e f_l f_i f_t L \quad , \qquad (1)$$

where $R_*$ represents the rate of star formation in the Galaxy, $f_p$ is the fraction of stars that host planets, $n_e$ is the mean number of planets in the so-called habitable zone³ of those stars with planets, $f_l$ is the fraction of those planets where life forms, $f_i$ is the fraction of life-bearing worlds that evolve intelligence, $f_t$ is the fraction of intelligent species that develop radio transmissions, and $L$ is the mean lifetime of such a transmitting technological species. In this paper we broaden the usual definitions of $f_t$ and $L$ beyond solely radio transmission to consideration of the emergence and longevity of *any* SWEIT, whether or not radio technology is involved.

Many analyses have been made attempting to produce estimates of *N* relevant to radio-based searches for other technological civilizations (Wallenhorst 1981, Pena-Cabrera & Durand-Manterola 2004). Although early work on this problem constituted a kind of educated guess work, those efforts were nevertheless useful in helping to structure debate about factors leading to the emergence of technological species. When Drake first proposed the equation only the first term $R_*$ could be estimated at all (the current best value is $R_* \approx 50$ stars/yr Prantozs 2013, Watson, private communication). In the last two decades, however, two more terms in the Drake Equation have become well-characterized. Beginning with the discovery in 1995 of an exoplanet orbiting the star 51 Peg, astronomers have now discovered roughly 1000 planets orbiting other stars (Howard 2013, Seager 2013). Most importantly, the current sample of exoplanets now allows a good estimate of the fraction of planet-bearing stars with estimates trending towards $f_p \sim 1$ (Seager 2013). In addition, studies of transiting planets, especially by NASA's Kepler mission, have also provided constraints on the number of exoplanetary *systems* (stars with multiple planets), their architecture, and the nature of the discovered planets. From this work the number of Earth sized worlds per system is estimated to be 14% (Dressing & Charbonneau 2013, Howard et al 2013). Grouping all terms dependent purely on astronomical investigations, i.e., $\dot{N}_a = R_* f_p n_e$, one estimates $\dot{N}_a = 7$ habitable planets formed per year, which implies that there now exist $\sim 10^{10}$ potentially habitable planets in the Galaxy.

---

³ In this paper we use the traditional definition of *habitable zone*: the range of orbital distances from its host star in which a planet's surface could have stable liquid water (taken as the *sine qua non* for life).

The next term in the Drake equation, $f_l$ (the fraction of habitable worlds in which life actually arises), remains completely unconstrained. There is, however, an expectation that over the coming decades estimates of its value may be possible. Studies of Mars in particular have yielded evidence for an early epoch in which water flowed on its surface. Future *in situ* explorations of Mars may provide evidence of fossil (or the lack thereof) for active life on the planet. In either case this will provide constraints on $f_l$. In addition, spectroscopic studies of exoplanet atmospheres also hold the possibility of yielding evidence for "biosignatures" in the form of non-equilibrium concentrations of atmospheric constituents that can be linked to an active biosphere (Seager 2013).

The possibility of empirically deriving constraints on Drake Equation factors ends, however, with the next two terms, $f_i$ and $f_t$. Both the fraction of life-bearing planets that evolve intelligence and the fraction of those that evolve technological cultures involve questions of evolutionary biology and sociology that are unlikely to be constrained without either *in situ* explorations of exoplanets (unlikely for centuries) or direct contact with a SWEIT. Even here, however, our knowledge of Earth's own evolutionary history allows some forms of inference to be attempted using probability theory (Lineweaver & Davis 2002, Spiegel & Turner 2012). Recognition that the evolutionary trajectory of life on our planet has passed through a number of *critical steps*, each of low probability, has allowed some authors such as Carter (1983, Watson 2008) to argue that probability distributions for $f_i$ and $f_t$, and hence, broad limits on their values, can be derived.

Consider, for example, that astronomical and geophysical factors give a planet a habitable lifetime of $t_h$. As shown in Carter (1983), if there are *n* critical steps leading to some property such as intelligence or technological capability and each k[th] step has low probability $\lambda_k$ ($\lambda_k t_h \ll 1$), then the expectation for the time at which the k[th] step will occur is

$$\langle t_{k/n} \rangle = \frac{k}{n+1} t_h \ . \qquad (2)$$

Using a more detailed model of the Earth's evolutionary history, Watson (2008) enlarged Carter's argument to estimate that there have been *n* = 7 critical steps for the emergence of intelligence. This implies that intelligence, on average, appears only at the very end of a planet's era of habitability. For stars like the sun this conclusion leads to a low value of both $f_i$ and $f_t$.

The discussion in these papers demonstrates the ways in which considerations of Earth's evolutionary history allow for broad astrobiological reasoning. For our purposes - exploring the relevance of astrobiology to key issues in sustainability - the debate over the correct value of the still unknown terms in the Drake Equation is less important than the existence of data and methods that advance that debate. In particular there is now a significant body of empirically derived knowledge about the planetary context of life either as it exists on Earth or its potential on other worlds. This may make it possible to address the single most important aspect of the Drake equation for sustainability science: *L,* the final factor. We therefore argue that a key question in sustainability science can be stated in explicitly astrobiological terms:

*Given an ensemble of N species with energy intensive technology (SWEITs), what is their average lifetime $\bar{L}$*

This is equivalent to asking: if we could rerun Earth's past (and future) history many times and select those trajectories leading to SWEIT (note they need not be human), then what value of $\bar{L}$ would be obtained across that ensemble of histories? While we do not know the actual value of $\bar{L}$, using the Drake Equation we can point to a methodology and start to answer this question.

First, we should check whether our proposed ensemble of SWEIT makes sense. We can calculate how large a volume of space is needed to contain a large enough SWEIT sample size *K* such that averages like $\bar{L}$ become meaningful. For *K*, we consider a value of 1000 to constitute a statistically relevant sample since the deviations around averages will, in a binomial distribution, go as $1/\sqrt{K}$, or just 3%. Since we are interested in the time-integrated number of SWEITs (meaning we include SWEITs already extinct), we can use an alternative form of the Drake Equation, considering it to be a probability distribution $dN_s$ of the number of SWEITs in a volume associated with a relevant cosmic scale (a galaxy, a cluster of galaxies, or observable Universe) relative to the lifetime $T_s$ of structures at that scale. Thus we write the Drake Equation as,

$$dN_s = N_{as} F_{bt} \, (dL/T_s), \qquad (3)$$

where $N_{as}$ is the total number of habitable planets in the volume associated with the scale considered ( $N_{as} = N_{*s} f_p n_e$ ), and $F_{bt} = f_l f_i f_t$ is the combination of biological and technological evolution factors which we call the "bio-technological probability". The total number of SWEITs (currently alive *or extinct*) at the chosen scale is the integral over all SWEIT lifetimes, or $N_s = N_{as} F_{bt}$ .

Assuming a constant density of stars (at whatever scale we are interested in), we can use this alternative form to derive a simple expression for what we call the *enclosure radius $R_e$* , which measures the size of a region containing a statistically relevant sample of technological species. Considering $n_s = N_s/(\frac{4}{3}\pi R_s^3)$ to be the number density of SWEIT over the given scale $R_s$ we have $\frac{4}{3}\pi R_e^3 n_s = K$ or,

$$R_e = R_s \left(\frac{K}{N_{as} F_{bt}}\right)^{1/3} \qquad (4)$$

For for galactic scales ($N_{as} \equiv N_{ag}$) we assume $f_p = 0.5, n_e = 0.5, N_{*g} = 10^{11}$, and $R_g \cong 10^5$ ly . This yields $N_{ag} \cong 10^{10}$ and $R_{eg} = 220$ lt_yr $(F_{bt})^{-1/3}$. For the scale of the observable Universe we assume the same values of $f_p$ and $n_e$, $R_u \cong 10^{10}$ ly and $N_{*g} = 10^{24}$. This yields $R_e = 2154$ lt_yr $(F_{bt})^{-1/3}$. In Fig 1 we plot the enclosure radius as a function of $F_{bt}$ .

Using these expressions we ask what is the minimum value of the bio-technological probability $F_{bt}^*$ that yields a statistically revelant sample in the local Univese $R_{lu}$. Taking $R_{lu} = 0.01 \, R_u$ we have $F_{bt}^* = 10^{-15}$. Thus even with the odds of evolving a SWEIT on a given habitable planet being one in one million billion at least 1000 species will still have passed through the transition humanity faces today within our local region of the cosmos.

Of course for less restrictive values of $F_{bt}$ a representative sample of species can be found within the local neighborhood of our Galaxy. For example assuming $F_{bt} = 10^{-6}$ yields $R_e = 3 \times 10^3$ ly.

This analysis demonstrates that from an astrobiological perspective the concept of a representative SWEIT ensample is reasonable even if the evolution of these species is highly unlikely. Over the course of cosmic evolution enough of SWEITs should have arisen to allow one, in principle, to meaningfully inquire about average properties of their developmental trajectories (such as $\bar{L}$). Note that we are explicitly *not* asking if any of these SWEIT could be contacted. They may already be long extinct. The exercise here is simply to determine if such an ensemble is meaningful to consider within a "local" volume of the Universe.

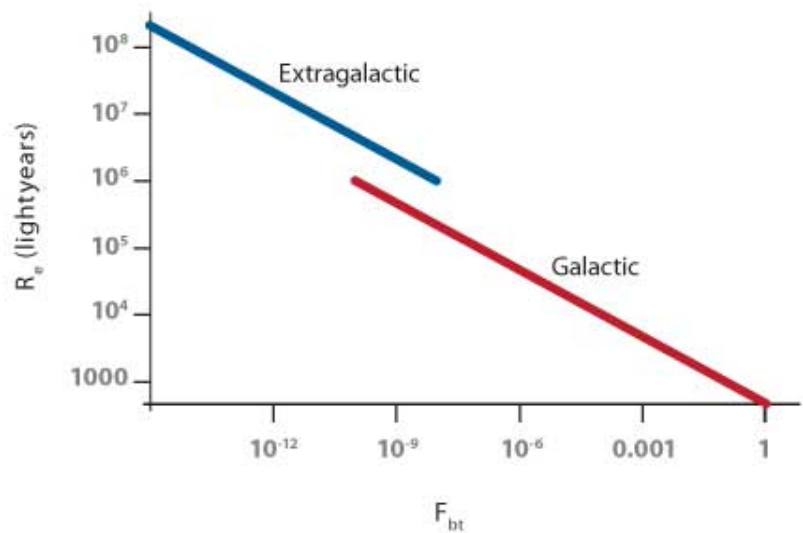

**Figure 1. Enclosure radius vs bio-technological probability. Radius of sphere $R_e$ enclosing a statistically significant ($K$ = 1000) SWEIT enseamble versus $F_{bt}$, the combined biological and evolutionary factors in the Drake equation. Blue line is appropriate for intergalactic distances (mean densities of stars smoothed over many galaxies) and is valid for distances above $R_e$ >$10^6$ light-years. Red line is appropriate for densities of stars within a galaxy and is valid for distances $R_e$ <$10^6$ light-years.**

Given the plausibility of the existence of a SWEIT ensemble, we now turn to exploration of how their trajectories of development can be conceptualized.

## 3. Trajectories of Technological Energy-Intensive Species

We now consider the general framework in which average "trajectories" for SWEIT evolution might be explored. We use the formalism of dynamical systems theory (Beltrami 1987) in which a set of governing differential equations is invoked to represent the evolution of a single SWEIT.

These equations, with appropriate initial conditions and constraints, determine the trajectory of the system through the multi-dimensional solution space defined by the system's independent variables. Invoking the concept of an ensemble of systems, each with different initial conditions and constraints, allows us to explore suitably averaged evolutionary characteristics such as a mean lifetime $\bar{L}$. The application of dynamical systems theory in this context is similar to that used in studies of systems ecology (May 1977, Petraitis & Hoffman 2010) and ecological economics (Krutilla & Reuveny 2006).

The SWEIT solution space is likely to be hyper-dimensional depending on conditions inherent to the planet on which the species evolved (ocean coverage, existence of plate tectonics, availability of various classes of resources such as fossil fuels, etc.) as well as biological and social characteristics (lifespans of individual organisms, degree of social cooperation, etc.). For our purposes in articulating a broad program of research, we consider a simplified set of equations and a concomitant solution space that may capture essential characteristics of the problem

As a toy model to illustrate the proposed method, consider an intelligent species - on the path to developing energy intensive technology - that can harvest some form of biomass for energy (such as trees in Earth's example). Thus the energy resource can be expressed as a "population" $E$. Let us assume that this renewable resource's own growth is limited by environmental factors which lead to a carrying capacity $K$. Thus the population of the energy bearing resource E is finite. The coupling between the growth of the SWEIT population $N$ and the energy resource population $E$ can then be described by a modified (logistic) form of a simple Predator-Prey system (Brauer & Castillo-Chavez 2011)

$$\frac{dE}{dt} = bE\left(1 - \frac{E}{K}\right) - aEN \qquad (5)$$

$$\frac{dN}{dt} = caEN - dN \qquad (6)$$

where $b$ is the growth rate of the energy resource, $a$ is the SWEIT "predation rate" of energy (its rate of energy harvesting), $c$ is the rate at which the resource can be used to increase the SWEIT population, and $d$ is the SWEIT mortality rate.

The behavior of such as system in its 2-D solution space $(N, E)$ is presents a textbook example of a stable dynamical system in which an initial population $N_o$ and resource $E_o$ experiences oscillatations with decreasing in amplitude until a steady state is achieved (at $N_s = b/a\,(1 - d/caK)$ and $E_s = d/ca$). Figure 2 shows a representative solution to the system given by equations 5 and 6.

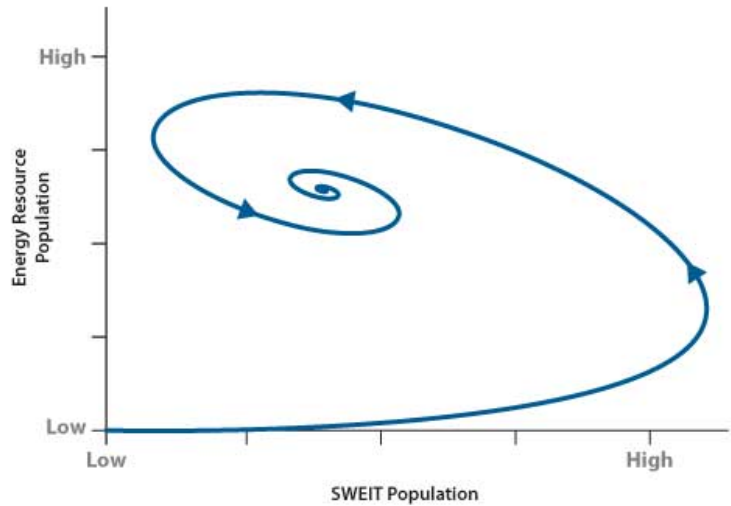

**Figure 2. Trajectory for coupled SWEIT/energy-resource populations described by modified predator-prey system (equations 5 & 6). The systems begins with low populations of both SWEIT and Energy resource $(N_o, E_o)$ and evolves to a stable solution as demonstrated by its trajectory approaching a fixed point $(N_s, E_s)$.**

The system of equations 5 and 6 represent a simple example of the idea of a solution space. In reality we would need a much more general description of SWEIT evolution. The lowest dimensional solution space appropriate for the evolution of a SWEIT would however have to include some form of feedback on the planetary system. Thus we need at least one independent variable that represents the "forcing" of the planetary systems by SWEIT activity. This forcing would likely represent, in some form, the energy released into the coupled planetary systems due to the technological use of harvested energy. Thus we define $r_f = D_t/D_n$ to be the ratio of the planetary system forcing driven by SWEIT growth ($D_t$) to that produced without the technological activity of the species (i.e. the planets own "natural" forcings, $D_n$). As an example consider that at the present epoch of human evolution the rate of energy trapping by anthropogenic $CO_2$ emission represents a form of $D_t$, while the rate of energy trapped as a result of volcanic outgassing of $CO_2$ represents a component of $D_n$.

Thus we suggest that a minimum solution space vector $\vec{S}$ for an individual SWEIT is $\vec{S} = (N, e_c, r_f)$ where $N$ is the SWEIT population, $e_c$ is the energy harvested per captia, and $r_f$ represents SWEIT feedback on planetary systems. Note that the variable $N$ captures the inherent success of the species defined from a purely Darwinian perspective; the variable $e_c$ captures the success of the species' technological capacities as it harvests and utilizes more energy than would be possible without technology; the variable $r_f$ captures reality of thermodynamic feedback such that energy used for work demands entropy generation. In particular, if $r_f \gg 1$ then the combined planetary systems will be driven into new states on timescales shorter than bio-evolutionary time scales for the SWEIT.

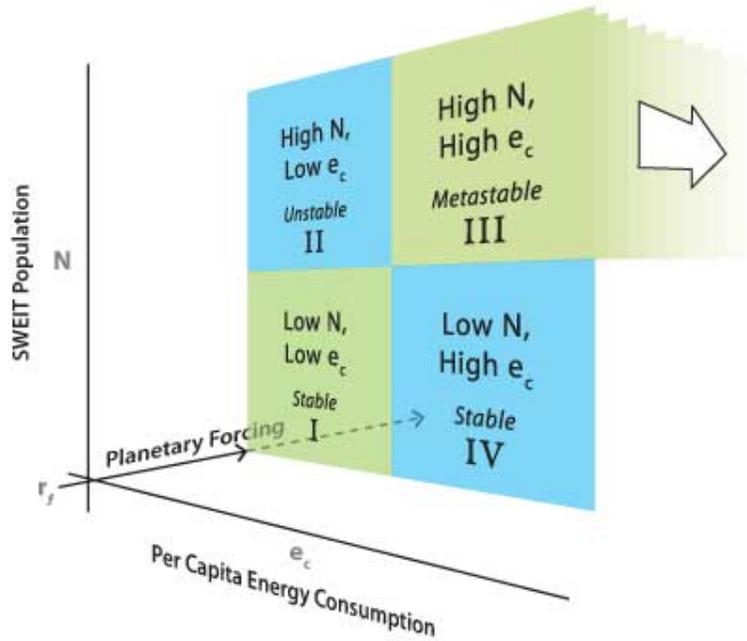

**Figure 3.** Solution space for SWEIT evolution. Shown is a schematic of possible stability domains in a single N (population), $e_c$ (energy harvest rate per capita) plane. As shown schematically in figure, region III is most likely to drive unstable increases in the direction of increased planetary forcing, $r_f$, which may degrade the planet's ability support a large SWEIT population.

A specific SWEIT evolutionary model takes the general form $\frac{dS}{dt} = \vec{M}(N, e_c, r_f, ...)$ where the vector of functions $\vec{M}(N, e_c, r_f, ...)$ reflects explicit couplings between terms. The research program is embodied in the exploration of different functional forms of $\vec{M}(N, e_c, r_f, ...)$. These different models should be informed by insights gained from astrobiological models (see Sec. 4). Those insights will also be expressed through the choice of initial conditions for the systems and additional constraints on variables or coupling constants between variables.

For a given specific model (i.e. a choice of $\vec{M}(N, e_c, r_f, ...)$) dynamical systems theory then allows analysis of regions of instability as well the presence of quasi-stable limit cycles or stable fixed points similar to what occurs in the simple Predator-Prey model presented above, (i.e. the stable fixed point at $(N_s, E_s)$ in Fig. 3). For a more complete model such global classes of behavior, i.e the existence of limit cycles, attractor basins, saddle points etc, should allow broad classes of behavior in SWEIT trajectories to be classified and understood.

Considering that the growth of the population *N* is likely to be coupled to the harvested energy (i.e., there is a positive feedback between the two variables) one should be able to map regions of stability and instability in the solution space. Figure 3 presents a schematic of possible regions of stability/instability in $(N, e_c)$ plane. Region I, with low population and low $e_c$, is likely stable in that a population could remain in that region for long periods relative to timescales inherent to the environment. Given the low energy harvesting capabilities this region would not be a SWEIT but something like human civilization many millennia ago. Region II, with high population and low $e_c$, would likely be unstable, as the SWEIT would consume the resources provided by the environment on timescales short compared with their natural regeneration time and would lack the energy harvest capacities to enhance the productivity of the environment. Region III, corresponding schematically to human society's current location, could in principle be stable as it maintains high populations through a large capacity to harvest energy. The difficulty for a SWEIT in this region, however, is that feedback on planetary systems generated as that energy is used will inevitably change $r_f$. This would mean that region III is unstable leading to movement perpendicular to the $(N, e_c)$ plane and eventually into regions where the environment is driven into new states on time scales short compared with natural responses. Finally, Region IV appears potentially stable (sustainable) as a smaller population with relatively high $e_c$ might not be capable of maintain stable values of $r_f$ over long time scales. Note that details of the locations of these quadrants will vary depending on the details of the system which, in turn, determine the limits at which instabilities set in. The purpose of an explicit modeling program would be to articulate such details including the size of the regions (shown schematically to be of equal area in Fig 3).

Figure 4 shows two classes of behavior in the full $(N, e_c, r_f)$ solution space. The red line represents a "collapse" trajectory whereby the species population initially increases along with its technology (energy harvesting capacities) at low $r_f$. Once a threshold of $N$ and $e_c$ is reached, however, entropy generation pushes the solution *rapidly* towards higher $r_f$ and global instability, thus leading to systemic failures of technological systems and negative feedbacks from the planetary systems on which those systems depend. Collapse of the population (low values of N) then follows. The blue line represents a sustainabile trajectory in which smaller levels of population with high technology are achieved before high values of $r_f$ negatively impact the ability of the species to maintain its energy harvesting systems. We represent the end state of such a solution as a limit cycle given the inherent time-dependence of such sustainable habitats (see Sec. 4.1). We expect that the topology of the solution space will dictate the actual form of these solutions in that both collapse and sustainable trajectories will be defined through attractor basins and unstable terrains such as saddle points between them (Brauer & Castillo-Chavez 2011).

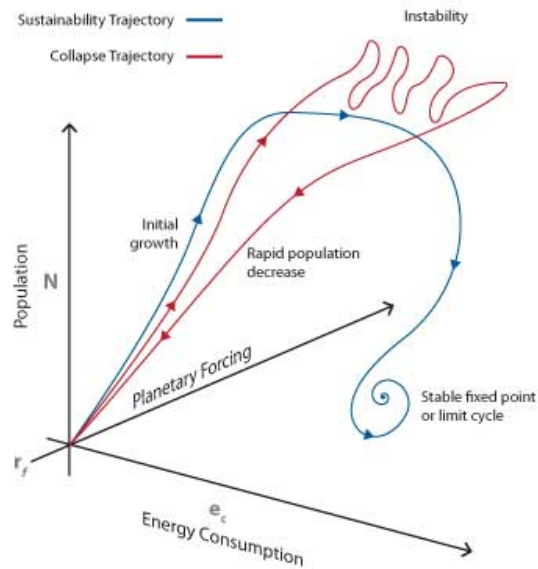

**Figure 4. Schematic of two classes of trajectories in SWEIT solution space. Red line shows a trajectory representing population collapse whereby development of energy harvesting technologies allows for rapid population growth which then drives increases in planetary forcing. As planetary support systems change state the SWEIT population is unable to maintain its own internal systems and collapses. Blue line shows a trajectory representing sustainability in which population levels and energy use approach levels that do not push planetary systems into unfavorable states.**

## 4. Areas of Astrobiology Relevant for Sustainability Science

In this section we outline three specific examples where astrobiological studies might drive new research aimed at developing a broader understanding of sustainable human cultures.

### 4.1 Planets, Moons and Exoplanets: Habitability, Sustainability and Time

A central concept in astrobiology is *habitability*, defined broadly as the ability of a planet to support life (Sullivan & Carney 2007); note that a planet may be habitable at a certain time even if no life actually exists. Habitability may also be restricted to include those conditions in which abiogenesis (the formation of life from non-life) can occur (Lineweaver & Chopra 2012), though this is not necessary because life could start elsewhere and be transported to a planet (panspermia) by a number of means (e.g., debris from planetary impacts occurring elsewhere in a planetary system).

Note that "Habitability" in astrobiology is a broader concept than "sustainability" in sustainability science. Instead of astrobiology asking *if any form of life is possible on an any given planet*, sustainability science asks *if, on a particular planet (Earth), a particular kind of human civilization (ours) is possible over long time scales*. Thus sustainability becomes a subset of habitability, albeit the subset with the greatest urgency to our society today. Given such a connection it is sensible to understand how progress in astrobiological studies of habitability can inform planetary perspectives on sustainability.

One of the most basic definitions of habitability corresponds to a planet having surface temperature ($T_p$) such that water molecules on the surface would be in a liquid state ($273 \text{ K} < T_p < 373 \text{ K}$ for a pressure of 1 atm). For a planet at a distance $D$ from a star with temperature $T_s$ and radius $R_s$ this yields,

$$T_p = (1-a)^{1/4} \frac{1}{2} \left(\frac{R_s}{D}\right)^{1/2} T_s \qquad (7)$$

where $a$ measures the planetary albedo (reflectivity) of incident starlight. The dependence of $T_p$ on distance means one can define a band of orbits (a range of $D$) called the Habitable Zone (HZ), where surface temperatures allow water to exist in the liquid state (Brownlee &

Kress 2007). $T_p$ (and thus the width and location of the HZ) can also be significantly altered by the radiative properties of the planet's atmosphere, e.g., $T_p$ is higher in the presence of greenhouse gases.

The discovery of subsurface "oceans" beneath the water-ice surfaced moons of giant planets in our solar system (e.g., Europa, Enceladus, and perhaps Titan) has enlarged our understanding and definition of habitability. Despite their huge distance from the sun, these moons maintain water in the liquid state due to tidal heating by their host planet (Chyba & Phillips 2007, Forgan et al 2013). Studies of Earth's oceans have also shown that purely chemotrophic ecosystems can develop around deep-sea hydrothermal vents without any need for energy from sunlight (Baross et al 2007). Thus it is possible that subsurface oceans on giant planet satellites might also be habitable (McKay 2007). The discovery of exoplanets in a variety of orbital configurations has also served to broaden definitions of habitability as astronomers consider a far wider range of orbital properties (e.g., rocky planets in tight orbits around cool dwarf stars) than those provided by our own solar system (Howard 2013). Thus the *spatial* domains of habitability have been significantly broadened through recent astrobiological studies.

For sustainability, however, a far more important domain may be astrobiological insights into the *temporal* domain of habitability. The capacity for a planet to support life is, fundamentally, a time-dependent property. This can be most easily seen in equation 7 where each factor in the equation can be expected to vary (though on different timescales). For example, the variation of the sun's temperature and radius ( $T_s, R_s$ ) has been moving its habitable zone outward over billion-year timescales. Based on solar evolution models it is estimated that increases in the sun's luminosity of ~10%/Gyr will render the Earth uninhabitable by most forms of present life as "soon" as ~1.0 Gyr in the future (Caldiera and Kasting 1991, Watson 2008, Kopparapu et al 2012)
.
The study of Mars provides the most direct and important example of time-dependent habitability. Studies by a variety of orbiters, landers and rovers over the last two decades have built a conclusive case that Mars once hosted a warmer, wetter climate (Azua-Bustos, et al 2012). In particular, early conditions on Mars may have allowed abiogenesis to occur (Jakosky et al 2007). The period of Martian habitability lasted less than 1 Gyr and ended perhaps ~4 Gyr ago. The cause of the loss of Martian habitability is still debated, but one favored theory invokes the early escape of its atmosphere's lighter elements and molecular species into space via collisions with the solar wind.

*One principal lesson to be drawn from astrobiological studies is that planetary habitability changes with time.* The same principle must surely apply to sustainability. Note that sustainability requires conditions appropriate not just for life but for a particular kind of social organization or a particular kind of species (a technological energy-intensive one). Thus naively one may expect that for cases like our own in which the appearance of strong SWEIT induced planetary forcing is just beginning, ($r_f \gg 1$), sustainable states may be inherently more delicately balanced . This would be the case because of competing forces where cultural dynamics are just as important as the response of the planet's biological and physical systems. Thus *we should expect that sustainability will be inherently easier to unbalance than just habitablity.* In other words one can expect the existence of multiple modes of instability amongst the coupled systems facing any SWEIT (Helbing 2012).

Modeling trajectories in systems coping with rapidly changing climate, for example, might draw on understandings derived from the history of Mars, Venus or Earth's own past, such as during the aftermath of mass extinctions (Sec. 4.2).

A more detailed understanding of habitability has already changed understandings of sustainability in terms of climate states. For example, studies of our own solar system (Kopparapu et al 2012) show that the Earth's orbit is now located just beyond the inner edge of the solar system's HZ (0.99 AU) . This means it is possible that our present climate is closer to the limiting case of a so-called "moist greenhouse" in which water vapor (an effective greenhouse gas) from ocean evaporation plays a significant role relative to $CO_2$ and other greenhouse gases. Had we found ourselves in the center of the Sun's HZ, the possibility of such a moist greenhouse would be more remote. It has been estimated that if we continue current anthropogenic $CO_2$ deposition rates into the atmosphere, Earth might enter the moist greenhouse phase by as early as the year 2300.

To summarize, studies of sustainability on a planetary scale can be seen as a subset of astrobiological concerns with habitability. Consideration of astrobiological studies of habitability demonstrate the inherent time dependence that equally applies to sustainability issues. Deeper consideration of astrobiological studies concerning the time dependence of habitability may thus prove also useful for sustainability studies.

**4.2 Mass Extinctions: Constraints on Responses to the Anthropocene**

Astrobiology is concerned with the long-term evolution of life on any planet where it might arise. The only example we have to date is, of course, that of Earth and so astrobiologists are deeply concerned with the "major" events in the history of Earth's life. Particular issues are: What constitutes a major event? When did these happen? What forces drove these events and would they be likely or inevitable for life on another planet (whether exotic life or our own form)? One particular class of major events includes *mass extinctions (*Ward 2007*)*.

The history of Earth's flora and fauna over the past 550 Myr provides evidence of five distinct mass extinctions. These were events in which the rate of extinction rose above the rate of speciation and more than 75% of all species were removed from the biosphere over a relatively short duration (Barnosky et al 2011). The "Big Five" mass extictions are: End-Cretaceous (KT) Event, 65 Ma (Myr ago) with an estimated 76% of all species lost; Triassic Event, 200 Ma, 80% lost; Permian Event, 251 Ma, 96% lost; Devonian Event, 359 Ma, 75% lost; and Ordovician Event, 443 Ma, 86% lost. In these events significant fractions of both land and marine species were rapidly driven into extinction, followed in each case by a significantly greater diversity of species. Furthermore, it has been argued that human activity is now driving the biosphere into a new, sixth mass extinction, the Anthropocene Event (Barnosky et al 2011). Thus mass extinctions represent another potential area in which astrobiological considerations can address both specific and foundational questions in sustainability science.

The best characterized mass extinction is the End-Cretaceous (KT) event which eliminated the dinosaurs and three-quarters of all other species, in the process eventually allowing mammals to gain a dominant foothold (Ward 2007). Most lines of evidence point to an asteroid or comet impact as the primary cause of this mass extinction. The KT event is, however, the only mass extinction that can definitely be associated with an impact.

The most significant mass extinction in Earth's history is currently ascribed to causes internal to earth systems coupling. The End-Permian Mass Extinction (251 Ma), involved the greatest loss of biodiversity in Earth's history. Within only 0.2 Myr, more than 56% of all genera and 96% of all species were lost. Current research links the cause of the Permian extinction event to large-scale volcanic magma flows associated with the formation of the Siberian Traps (Payne and Clapham 2012). The release of $CO_2$ from the flows triggered enhanced global warming which then initiated a strong response from the climate system. Changes in ocean levels of $CO_2$ (and therefore acidity) and subsequent changes in ocean circulation led to deep marine anoxia. The alterations in ocean circulation also led to changes in ocean stratification. Purple algae brought into contact with sunlight-rich layers may then have driven production of high levels of hydrogen sulfide both at sea and over land, further enhancing extinction levels.

Thus it appears that climate change, driven by enhanced release of greenhouse gases, was the agent driving the most powerful mass extinction in Earth's history. In fact, other than the KT event, it may well be that climate change driven by increased greenhouse gas concentrations through volcanism played a major role in all mass extinction events (Payne and Clapham 2012, Fuelner 2008). The implication of rapidly changing climate as either a direct or secondary cause of previous mass extinctions holds obvious lessons for our own situation (i.e. environmental stresses driving cascades of extiction, Newman 1997). Understanding the ways climate has coupled to significant changes in biodiversity in the past is one direct application of astrobiology to sustainability studies. This is particularly true as such understanding can be applied to current conditions in which anthropogenic climate change and other anthropogenic drivers are forcing the coupled earth systems.

The record of previous mass extinctions has other uses for sustainability studies, including, for example, characterization of 21st century trends in biodiversity. Consideration of the fossil record, for example, leads to estimates that the time to a 75% reduction in species (i.e., an "official" sixth mass extinction) will be just 200-600 yr from now (assuming that current loss rates in biodiversity hold, Barnosky et al 2011). Such extrapolations into the future remain uncertain, but can be improved by looking both backward, in terms of better accounts of previous extinction events, and forward, in terms of better modeling of the impact of the Anthropocene on the biosphere. Such modeling can be done within the general context of SWEIT trajectory determination in order to set limits on the sensitivity of a SWEIT to rapid changes in biodiversity .

**4.3 The Great Oxidation Event: Predicting Climate Change For SWEIT**

Another series of key events in the history of Earth's life involved profound changes in the composition of the atmosphere. In particular the concentration of oxygen in both the early Earth's atmosphere and oceans was far lower, with atmospheric values probably $< 10^{-5}$ of current values (Catling & Kasting 2007). The emergence of an oxygen rich atmosphere was a key event in the history of the planet (Kasting & Kirschvink 2012). Furthermore, it was an event driven by activity within the biosphere itself through respiration by phototrophic prokaryotic single-celled organisms (anaerobic photosynthesis REF). The most dramatic increase in oxygen levels, by a factor of at least $10^4$, occurred relatively rapidly 2.4 Ga in the so-called Great Oxidation Event (GOE) The GOE (and secondary later increases) represents one of the most important examples of co-evolution between life and the planet and, as such, has important implications for astrobiology and sustainability (Arndt & Nisbit 2012).

While the ultimate source of increasing $O_2$ levels is universally recognized to have originated in the biosphere (i.e. non-oxygen based photosynthesis), the reasons remain debated for the relatively rapid (on geologic timescales) increase in atmospheric $O_2$ levels in the GOE. Proposed causes include (Catling & Kasting 2007): changes in the chemical state of outgassed mantle material; a decrease in atmospheric methane levels (Kasting et al 1983, Pavlov 2000); changes in ocean sinks for $O_2$. While significant uncertainty remains as to *why* the $O_2$ levels increased when they did, in all scenarios strong effects on the coupled Earth systems occur. For example, an anoxic atmosphere would likely have had high concentrations of methane, a gas that on a molecule-by-molecule basis is ~25 times more potent as a greenhouse absorber than $CO_2$. Increases in $O_2$ would remove methane from the atmosphere, decreasing average global temperatures. The possible occurrence of several "Earth" episodes (near total glacial coverage) occuring during or just after the GOE may also be related to this removal of methane (Kirschvink et al 2000).

Just as important was the biosphere's own reaction to the increase in $O_2$ levels. The GOE was a double-edged sword for life as the energetics of oxygen chemistry are destructive for biological activity, yet also allow for more efficient metabolic pathways (Leigh,et al 2007). Aerobic respiration, for example, is ~16 times more efficient than anaerobic processes at generating ATP, the primary energy carrier for metabolism (Gaidos & Knoll 2012). Thus those species that did not evolve a means for oxygen detoxification became trapped in anaerobic niches, while the remainder evolved new behaviors that allowed them to thrive in the newly oxygen-rich atmosphere and oceans.

The GOE has significant implications for the astrobiological perspective on sustainability. In particular, it shows that in at least one instance life has significantly altered the chemical composition of a planet's atmosphere. From this vantage point, the current era of anthropogenic climate change can be seen more broadly than as an anomalous byproduct of human technological progress.

From the GOE we see that at least one time in the past life strongly forced the coupled earth systems. In light of this the alteration of atmospheric chemistry *might be expected* to occur as a consequence of rapid technologically-driven energy harvesting within some subset of SWEIT trajectories. Capturing this feedback would be part of any program studying such trajectories even in restricted models considering, say, only coupling between $r_f$ and $e_c$, i. e.,

$$\frac{dr_f}{dt} \propto f(e_c, \frac{de_c}{dt}, \ldots) ,$$

where $f(e_c, \frac{de_c}{dt}, \ldots)$ models the links between energy harvesting, greenhouse gas production and and other quantities which drive forcing changes in $r_f$.

Consider for example trajectory of our own species over the last 10,000 years of its evolution. Fig 5a tracks the human population, total energy consumption and as a proxy for planetary forcing, the atmospheric abundance of carbon dioxide ($CO_2$). The plot shows the rapid *coupled-increase* in all 3 quantities within the last century. In Fig 5b we present the same data as a trajectory in the SWEIT solution space introduced in section 3. The rapid coupled increase in population, energy consumption and planetary forcing mirrors the initial phases of the schematic trajectories shown in Fig 4. The question we hope our approach might address would be to explicate the various forms of behavior that can expected in the future: sustainability, collapse or some middle ground. As we have demonstrated the astrobiological perspective allows the broad considerations of links between planetary systems and biospheric activity (in this case due to a single SWEIT species) to be articulated and, hopefully, modeled.

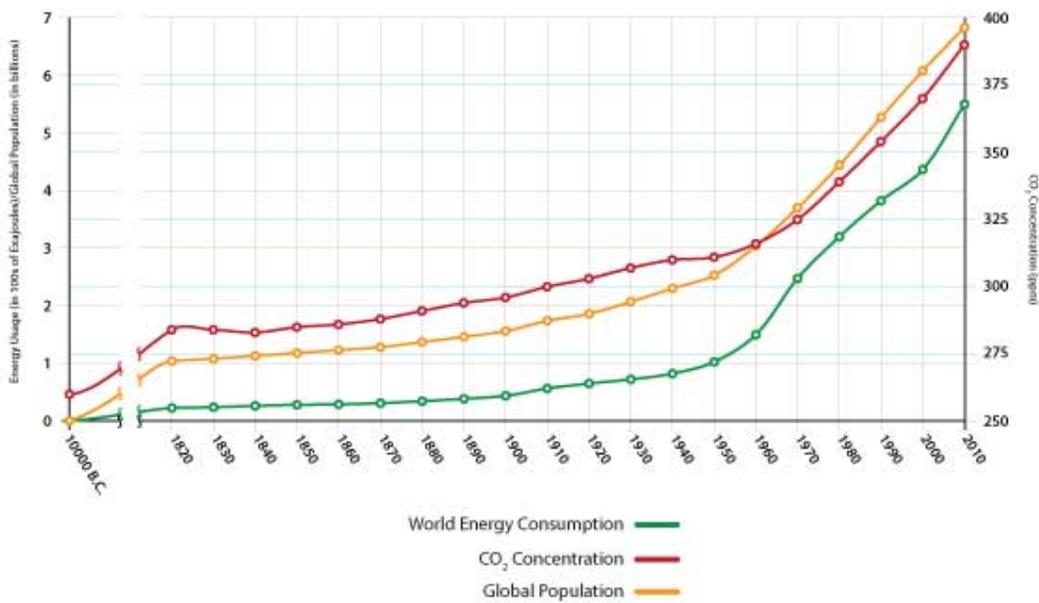

**Figure 5a. Plot of human population, total energy consumption and atmospheric $CO_2$ concentration from 10,000 BCE to today. Note the *coupled increase* in all 3 quantities over the last century.**

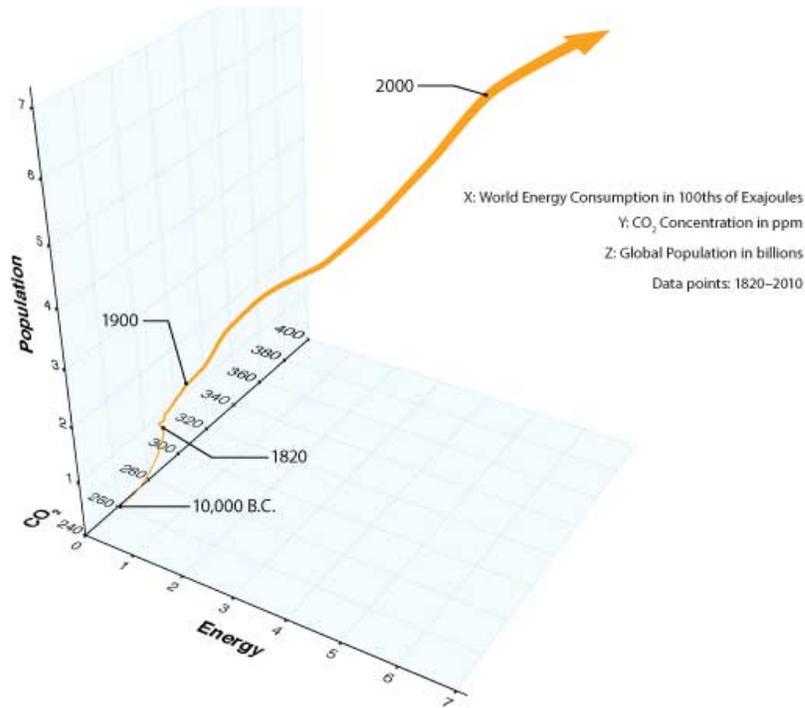

**Figure 5b. Same data as figure 5a but plotted as trajectory in SWEIT solution space**

Consideration of Figures 5a and 5b indicates that modeling the trajectory of SWEIT ensembles would necessarily include the evolution of atmospheric changes. One outcome of such modeling may be to find that a greenhouse phases and a sustainability "crisis" be a generic feature of at least some subset of the ensemble. The question then becomes how common and enduring such a phase is, and, more importantly, what characterizes successful paths *out of* such a crisis, i.e., what leads to long mean SWEIT lifetimes $\bar{L}$?

## 5. Discussion and Summary

Sustainability science and astrobiology both seek to understand the intimate, symbiotic, and continually evolving connections between life and host planets. Sustainability science is focused on the effects of one particular species during one particular epoch, whereas astrobiology broadens its purview to all possible species on Earth or elsewhere. Both fields are rapidly changing and in their infancy. Astrobiology's concept of planetary habitability is of particular relevance to sustainability science, while the latter's concern with rapid changes in the biosphere caused by a single intelligent species (ours) informs those astrobiologists considering the possible existence of Species with Energy-Intensive Technology (SWEITs) on other planets.

In this paper we have suggested the beginnings of a research program that we hope will benefit both sustainability science and astrobiology, emphasizing in particular how recent developments in astrobiology are of direct relevance to sustainability science. One promising avenue comes through considering an ensemble of SWEITs that exist now or in the past (located well outside of our solar system and perhaps even our Galaxy). Each SWEIT's history defines a trajectory in a multi-dimensional solution space with axes representing quantities such as energy consumption rates, population and planetary systems forcing from causes both "natural" and driven by the SWEIT itself. Using dynamical systems theory, these trajectories can be mathematically modeled in order to understand, on the astrobiology side, the histories and mean properties of the ensemble of SWEITs, as well as, on the sustainability science side, our own options today to achieve a viable and desirable future. We note that dynamical systems theory as we have presented it represents only one theoretical methodology possible. For example, the use of network theory with its emphasis on multiple cascading paths to system failures (or system resiliency) may also prove useful (Helping 2012).

Modeling SWEIT/planet evolution in the way we have described may allow for broad classes of behavior to be articulated. A future research project may, for example, explore if the development of enhanced greenhouse forcing is an *expected* outcome of SWEIT evolution. This could occur based both on the most likely energy sources harvested early in a species' technological development and/or planetary changes driven a results of other SWEIT activity.

We note in conclusion that in Payne & Clapham's (2012) review of the End-Permian mass extinction (subtitled "An Ancient Analog for the 21st Century?") the authors state: "The geological record is increasingly essential as an archive of past experiments in global change… The best - and most sobering - analogs for our near future may lie deeper in Earth's past." The purpose of this paper has been to broaden such a conclusion to include the whole of astrobiological studies. Recent studies exploring planetary-scale tipping points (Lenton & Williams 2013) or the existence of planetary-scale "boundares" as safe operating limits for civilization (Rockstrom et al 2009) emphasize that a astrobiological perspective is already present, if not fully recognized in many modern sustainability studies.

Thus the evidence is indeed strong that during our present Anthropocene epoch the coupled Earth systems are being altered on an extremely rapid time scale. Although such rapid changes are not a new phenomenon, the present instance is the first (we know of) where the primary agent of causation is knowingly watching it all happen and pondering options for its own future. In this paper we have argued that it is unlikely that this is the first time such an event as occurred in cosmic or even galactic history. The point is to see that our current situation may, in some sense, be natural or at least a natural and generic consequence of certain evolutionary pathways. Given that fact it may be possible to use the data and perspectives of astrobiology to tell us something about optimal pathways forward. One point is clear, both astrobiology and sustainability science tell us that the Earth will be fine in the long run. The prospects are, however, less clear for *Homo sapiens.*

**Acknowledgements:** We wish to thank Bruce Balick, Don Brownlee, David Catling, Dan Watson, Marcelo Glieser, John Tarduno, Marina Alberti and James Kasting for helpful discussions. Eric Blackman provided helpful comments on an early draft. This work was supported in part with funds from the University of Rochester.

**Bibliography**

Alberti, M. Russo, Tenneson. (2013). *Snohomish Basin 2060 Scenarios: Adapting to an Uncertain Future.* Urban Ecology Research Laboratory, University of Washington, Seattle.

Arndt, Nicholas T., Nisbet, Euan G. (2012). "Processes on the Young Earth and the Habitats of Early Life." *Annual Review of Earth and Planetary Sciences,* vol 40, pp. 521–49.

Arnould, Jacques. (2009). "Astrobiology, Sustainability and Ethical Perspectives". *Sustainability,* vol. 1, pp 1323 – 1330.

Azua-Bustos, A., Vicuna, R., & Pierrehumbert, R., 2012, "Early Mars", in Frontiers in AstroBiology, Chris Impey, Jonathan Lunine, Joses Funes Eds, Cambridge University Press, Cambridge, United Kingdom and New York, NY, USA, 157-175

Barnosky, Anthony D., *et al.* (2011). "Has the Earth's sixth mass extinction already arrived?" Macmillan Publishers Limited, *Nature,* vol. 471, pp 51-57.

Baum, Seth D. (2010). "Is Humanity Doomed? Insights from Astrobiology." *Sustainability,* vol. 2, pp 591-603.

Barreiro, Marcelo, Fedorov, Alexey, Pacanowski, Ronald, Philander S. George. (2007). "Abrupt Climate Changes: How Freshening of the Northern Atlantic Affects the Thermohaline and Wind-Driven Oceanic Circulations." *Annual Review of Earth and Planetary Sciences,* vol. 36, pp. 33–58

Blois, Jessica L., Hadly, Elizabeth A. (2009). "Mammalian Response to Cenozoic Climatic Change." *Annual Review of Earth and Planetary Sciences,* vol. 37, pp. 181–208.

Beltrami, Edward, (1986), "Mathematics for Dynamic Modeling", Academic Press, San Diego, CA.

Brauer & Castillo-Chavez, (2011), "Mathematical Models in Population Biology and Epidemiology", Springer, New York

Caldeira, K. and Karting, J.F. (1992), "The life-span of the biosphere revisited". *Nature* 360, 721–723.

Carter, B., McCrea, W. H. (1983). *The Anthropic Principle and its Implications for Biological Evolution.* Royal Society Publishing, pp 347-363.

Castilo-Rogez, J., & Lunine, J., 2012, "Small Habitable Worlds" in *Frontiers in AstroBiology*, Chris Impey, Jonathan Lunine, Joses Funes Eds, Cambridge University Press, Cambridge, United Kingdom and New York, NY, USA, 201-228

Costanza, Robert, *et al.* (2007). *Sustainability or Collapse: What Can We Learn from Integrating the History of Humans and the Rest of Nature?* Royal Swedish Academy of Sciences, pp 522-527.

Coustenis, A., & Blanc, M., 2012, "Large Habitable Moons", in *Frontiers in AstroBiology*, Chris Impey, Jonathan Lunine, Joses Funes Eds, Cambridge University Press, Cambridge, United Kingdom and New York, NY, USA, 175-201

Decker, Ethan H., *et al*. (2000). *Energy and Material Flow Through the Urban Ecosystem.* Annual Reviews, pp 685-740.
www.annualreviews.org

Dressing C., & Charbonneau, D., 2013, *The Occurence Rate of Small Planets around Small Stars", ApJ, 767, 95*

Feulner, Georg. (2009). *Climate Modelling of Mass–Extinction Events:A Review.* Paper for the ESLAB 2008 Special Issue of IJA, Postdam, Germany.

Forgan, Duncan, Kipping, David. (2013). *Dynamical Effects on the Habitable Zone for Earth-like Exomoons.* Cornell University Library, Cornell, NY. arXiv:1304.4377.
http://lanl.arxiv.org/abs/1304.4377v1

Goldblatt, Colin, Watson, Andrew J. (2012). *The Runaway Greenhouse: implications for future climate change, geoengineering and planetary atmosphere.* Cornell University Library, Cornell, NY. arXiv:1201.1593.
http://arxiv.org/abs/1201.1593v1

Holling, C. S. (2001). "Understanding the Complexity of Economic, Ecological, and Social Systems." *Ecosystems,* vol. 4, pp 390-405.

Howard, A.W., (2013) Observed Properties of Extrasolar Planets. Science 340, 572-576.

Howard, A., et al 2013, P*lanet Occurrence with 0.25 AU of Solar-type stars from Kepler*, ApJ, 201, 15

Helbing, D., 2012, "Globally networked risks and how to respond", Nature 497, 51–59

Kane, Stephen R., Hinkel, Natalie R. *On the Habitable Zones of Circumbinary Planetary Systems.* Cornell University Library, Cornell, NY., arXiv:1211.2812.
http://arxiv.org/abs/1211.2812v1

Kates, Robert W. (2010). *Readings in Sustainability Science and Technology.* CID Working Paper No. 213, Harvard College, Cambridge, MA.

Kates, Robert W. (2011). "What kind of a science is sustainability science?" *Proceedings of the National Academy of Science of the United States of America*. vol. 108, no. 49, pp. 19449–19450.

http://www.pnas.org/content/108/49/19449

Kates, Robert W. (2011). *From the Unity of Nature to Sustainability Science: Ideas and Practice.* CID Working Paper No. 218, Harvard College, Cambridge, MA.

Kasting, J., & Kirschvink, J., 2012, "Evolution of a habitable planet Evolution of a habitable planet" in Frontiers in AstroBiology, Chris , Jonathan Lunine, Joses Funes Eds, Cambridge University Press, Cambridge, United Kingdom and New York, NY, USA, 115-132

Kleidon, Axel. (2011). "Life, hierarchy, and the thermodynamic machinery of planet Earth." *Elsevier*, Vol. 7, Issue 4, pp 424–460.
http://www.sciencedirect.com/science/article/pii/S1571064510001107

Kleidon, Axel. (2012). *How does the Earth system generate and maintain thermodynamic disequilibrium and what does it imply for the future of the planet?* Royal Swedish Academy of Sciences, pp 1012-1040.

Kopparapu, Ravi kumar, *et al.* (2013). *Habitable Zones Around Main-Sequence Stars: New Estimates.* Cornell University Library, Cornell, NY., arXiv:1301.6674.
http://arxiv.org/abs/1301.6674v2

Krutilla, Kerry, Reuveny, Rafael. (2006). "The systems dynamics of endogenous population growth in a
renewable resource-based growth model." *Elsevier,* vol. 56, pp 256– 267.

Lenton, T.M & Williams, H.T.P., (2013), "The Origin of Planetary Scale Tipping Points", Trends in Ecology & Evolution, 28, 380-382

Lineweaver, Charles H., Chopra, Aditya. (2012). "The Habitability of Our Earth and Other Earths: Astrophysical, Geochemical, Geophysical, and Biological Limits on Planet Habitability."  *Annual Review of Earth and Planetary Sciences*, vol. 40, pp. 597–623.

Liu, Jianguo, *et al.* (2007). "Complexity of Coupled Human and Natural Systems." *Science*, vol. 317, pp. 1513-1516.

Loarie, Scott R., *et al.* (2009). "The velocity of climate change." Macmillan Publishers Limited, *Nature*, vol. 462, pp 1052-1055.

May, Robert M. (1977). "Thresholds and breakpoints in ecosystems with a multiplicity of stable states." Nature Publishing Group, *Nature,* vol. 269, pp. 471-477.

Mayor, M. et al. (2011), "The HARPS search for southern extra-solar planets XXXIV. Occurrence, mass distribution and orbital properties of super-Earths and Neptune-mass planets", e-print arXiv:1109.2497

Marshall, Charles R. (2006).  "Explaining the Cambrian "Explosion" of Animals."  *Annual Review of Earth and Planetary Sciences,* vol. 34, pp. 355–84.

Mann, Michael E. (2007). "Climate Over the Past Two Millennia." *Annual Review of Earth and Planetary Sciences,* vol. 35, pp. 111–36.

McInerney, Francesca A., Wing, Scott L. (2011). "The Paleocene-Eocene Thermal Maximum: A Perturbation of Carbon Cycle, Climate, and Biosphere with Implications for the Future." *Annual Review of Earth and Planetary Sciences,* vol. 39, pp 489–516.

Naganuma, Takeshi. (2009). "An Astrobiological View on Sustainable Life." *Sustainability,* vol. 1, pp. 827-837.

Newman, M. E. J. (1997). "A Model of Mass Extinction." *Science Direct,* vol. 189, pp. 235-252.

Petraitis, Peter S., Hoffman, Catharine. (2010). "Multiple stable states and relationship between thresholds in processes and states." *Marine Ecology Progress Series,* vol. 413, pp. 189-200.

Pena-Cabrera, G.V.Y., Durand-Manterola, H.J.(2004),Number of Planets with life in the galactic habitable zone deduced by modified Drake Equation, 35th COSPAR Scientific Assembly 35, 1903.

Payne, Jonathan L., Clapham, Matthew E. (2012). "End-Permian Mass Extinction in the Oceans: An Ancient Analog for the Twenty-First Century?" *Annual Review of Earth and Planetary Sciences,* vol. 40, pp. 89–111.

Prantzos, Nikos*.* (2013). *"*A Joint Analysis of the Drake equation and the Fermi Paradox", *International Journal of Astrobiology, 12, 246-253*

Roe, Gerard. (2009). "Feedbacks, Timescales, and Seeing Red." *Annual Review of Earth and Planetary Sciences*, vol. 37, pp. 93–115.

Rockstrom, J., et al (2009), "Planetary Boundaries: Exploring the Safe Operating Space for Humanity", Ecology and Society, 14, 32-55

Seager, S., (2013) ExoPlanet Habitability. Science 340, 577-581.

Seager, S, 2012, "Searches for Habitable Planets" in Frontiers in AstroBiology, Chris Impey, Jonathan Lunine, Joses Funes Eds, Cambridge University Press, Cambridge, United Kingdom and New York, NY, USA

Spiegel, D & Turner, E., (2012), "Bayesian analysis of the astrobiological implications of life's early emergence on Earth", PNAS, 109, 395

Smith, David A. (1977). "Human Population Growth: Stability or Explosion?" *Mathematics Magazine*, Vol. 50, pp. 186-197.

Sullivan & Baross 2007, "*Planets and Life"*, eds. W. T. Sullivan, III , J. Baross, Cambridge, UK: Cambridge University Press.

Sullivan & Carney 2007, "The History of Astrobilogical Ideas" in *Planets and Life*, eds. W. T. Sullivan, III , J. Baross, Cambridge, UK: Cambridge University Press, 7-45

Tattersall, Ian, Schwartz, Jeffrey H. (2009). "Evolution of the Genus Homo." *Annual Review of Earth and Planetary Sciences*, vol. 37, pp. 67–92.

Tarter, J., (2007), Searching for Extraterrestrial Intelligence. In *Planets and Life*, eds. W. T. Sullivan, III , J. Baross, Cambridge, UK: Cambridge University Press.

"Sustainability and the Ehrlich equation." *Population Matters.* (2011). http://www.populationmatters.org/wp-content/uploads/ipat.pdf.

Wiley, Keith B. (2011). *The Fermi Paradox, Self-Replicating Probes, and the Interstellar Transportation Bandwidth.* Cornell University Library, Cornell, NY., arXiv:1111.6131. http://arxiv.org/abs/1111.6131

Watson, Andrew J. (2008). "Implications of an Anthropic Model of Evolution for Emergence of Complex Life and Intelligence." *Astrobiology,* vol. 8, pp. 1-11.

Winguth, Arne Max Erich. (2011). *The Paleocene-Eocene Thermal Maximum: Feedbacks Between Climate Change and Biogeochemical Cycle.* The University of Texas at Arlington, Arlington, TX.

Ward, P, (2007), "Mass Extictions", In *Planets and Life*, eds. W. T. Sullivan, III , J. Baross, Cambridge, UK: Cambridge University Press.

Wallenhorst, S. G.,(1981), The Drake Equation Reexamined. The Quarterly Journal of the Royal Astronomical Society 22, 380.